\documentclass[11pt]{article}
\usepackage[margin=1in]{geometry}
\usepackage{amsmath,amssymb,amsthm}
\usepackage{fontspec}
\usepackage{fancyvrb}
\usepackage{booktabs}
\usepackage{xcolor}
\usepackage[hidelinks]{hyperref}
\usepackage{url}
\fvset{frame=single,framerule=0.3pt,fontsize=\small,xleftmargin=1em,xrightmargin=1em}

\newtheorem{theorem}{Theorem}
\newtheorem{proposition}[theorem]{Proposition}
\newtheorem{lemma}[theorem]{Lemma}
\theoremstyle{definition}
\newtheorem{definition}[theorem]{Definition}
\theoremstyle{remark}

\newcommand{\lean}[1]{\texttt{#1}}
\newcommand{\DTW}{\mathrm{DTW}}
\title{A Machine-Checked Theory of Dynamic Time Warping in Lean~4}
\author{Srihari Maruthachalam\\ \small Independent researcher\\
  \small \href{mailto:srihari.m.srihari@gmail.com}{\texttt{srihari.m.srihari@gmail.com}},
  \href{mailto:harimaruthachal@alumni.iitm.ac.in}{\texttt{harimaruthachal@alumni.iitm.ac.in}}}
\date{}

\begin{document}
\maketitle

\begin{abstract}
Dynamic time warping (DTW) is a standard dissimilarity measure for time series. In practice,
large-scale DTW search depends on two ingredients whose correctness is usually argued on paper:
global path constraints such as the Sakoe--Chiba band, and cheap lower bounds such as LB\_Kim and
LB\_Keogh that prune candidates without false dismissals. We present a formalization of DTW in
Lean~4 with mathlib, covering:
warping paths and DTW as a minimum over them;
correctness of the dynamic-programming recurrence;
nonnegativity, symmetry and $\DTW(x,x)=0$, with a machine-checked witness of the known failure of
the triangle inequality;
constrained and Sakoe--Chiba banded DTW;
the LB\_Kim, LB\_KimFL, LB\_Yi and LB\_Keogh lower bounds, the last restated for an arbitrary global
constraint;
and an executable row-by-row implementation proved equal to the specification.
The recurrence is proved once, for cell costs in an arbitrary linearly ordered additive commutative
monoid. Instantiating it at $\mathbb{R}\cup\{\top\}$ yields constrained DTW with no separate
optimality proof.
The development has no \lean{sorry}, and every main theorem uses only Lean's standard axioms.
To our knowledge it is the first formalization of DTW itself in a proof assistant. The mathematics is
classical; the contribution is a checked, reusable development and the precision it forces.
\end{abstract}

\section{Introduction}

DTW aligns two sequences by a monotone, continuous warping of their time axes and charges the sum
of local costs along the alignment~\cite{sakoe1978dynamic,keogh2005exact,muller2007dtw}. Its quadratic cost
motivated a literature of global constraints and lower bounds. The Sakoe--Chiba
band~\cite{sakoe1978dynamic} restricts the warping path to a diagonal band. LB\_Kim~\cite{kim2001index}
and LB\_Keogh~\cite{keogh2005exact} are linear-time lower bounds that let a nearest-neighbour search
skip the full DTW computation without false dismissals, and they are central to systems such as the
UCR suite~\cite{rakthanmanon2012searching}. The correctness of this pruning rests on the
lower-bound inequalities, which are proved informally.
DTW originated in speech recognition~\cite{sakoe1978dynamic,itakura1975minimum} and is used well
beyond it, for example to classify artifacts in EEG
signals~\cite{maruthachalam2019time,maruthachalam2020eeg}.

We formalize this theory in Lean~4~\cite{moura2021lean4} on top of mathlib~\cite{mathlib2020}.
Our contributions are:
\begin{itemize}
  \item \textbf{A generic optimality theorem} (Section~\ref{sec:dp}): for any cell cost
  $C:\mathbb{N}\to\mathbb{N}\to\beta$ into a linearly ordered additive commutative monoid
  $\beta$, the DP table equals the least cost of a warping path, and the least cost is attained.
  No sign assumption is needed.
  \item \textbf{Basic properties} (Section~\ref{sec:props}): nonnegativity, symmetry and
  $\DTW(x,x)=0$, each under the hypotheses its proof uses. Also a witness, for equal-length
  sequences and the cost $|a-b|$, of the well-known failure of the triangle
  inequality~\cite{muller2007dtw,lemire2009faster}, with the DTW values computed in Lean from the
  definition.
  \item \textbf{Constraints} (Section~\ref{sec:band}): constrained DTW as the same DP over
  $\mathbb{R}\cup\{\top\}$. Results: monotonicity in the admissible region, so banded DTW is
  non-increasing in the band width and bounded below by unconstrained DTW; exactness for wide bands;
  and finiteness exactly when $|m-n|\le w$.
  \item \textbf{Lower bounds} (Section~\ref{sec:lb}): LB\_KimFL for any nonnegative cost when
  $(m,n)\neq(0,0)$, and LB\_Kim and LB\_Keogh for every cost $f(|a-b|)$ with $f$ monotone and
  nonnegative on $[0,\infty)$. This covers both $|a-b|$ and $(a-b)^2$. LB\_Keogh is restated, with
  the proof of~\cite{keogh2005exact}, for an arbitrary global constraint and unequal lengths. The
  constant-width case of their Proposition~1 and LB\_Yi are corollaries.
  \item \textbf{An executable DP} (Section~\ref{sec:fast}) that keeps a single row, proved equal to
  the specification and cross-checked against an independent implementation.
\end{itemize}
Formalizing also made two small points about the presentation in~\cite{keogh2005exact} explicit
(Section~\ref{sec:props}). Their path conditions, read literally, admit repeated cells, which the
step-size definition of M\"uller~\cite[Def.~4.1]{muller2007dtw} excludes. And the path-length bound
stated there and in~\cite[Eq.~(3)]{ratanamahatana2004making} holds with $\le$, not $<$.

\section{Definitions}\label{sec:defs}

We index from $0$. A sequence with last index $m$ is a function $x:\mathbb{N}\to\alpha$ of which only
$x_0,\dots,x_m$ matter. This avoids dependent \lean{Fin} arithmetic. Lemma~\ref{lem:bounds} shows that
warping paths stay in the rectangle $[0,m]\times[0,n]$, and \lean{table\_congr} that the DP table reads
no cost outside it.

\begin{definition}
A \emph{step} from cell $r$ to cell $q$ is one of the moves $(1,0)$, $(0,1)$, $(1,1)$. A \emph{warping
path} to $(m,n)$ is a list of cells that starts at $(0,0)$, ends at $(m,n)$, and whose consecutive
cells are related by a step.
\end{definition}

\begin{Verbatim}
def Step (p q : ℕ × ℕ) : Prop :=
  (q.1 = p.1 + 1 ∧ q.2 = p.2) ∨ (q.1 = p.1 ∧ q.2 = p.2 + 1) ∨
    (q.1 = p.1 + 1 ∧ q.2 = p.2 + 1)

structure IsWarpingPath (m n : ℕ) (p : List (ℕ × ℕ)) : Prop where
  head  : p.head? = some (0, 0)
  last  : p.getLast? = some (m, n)
  chain : p.IsChain Step
\end{Verbatim}

The cost of a path under a cell cost $C$ is $\sum_{(i,j)\in p} C(i,j)$. The DP table is defined by
well-founded recursion:
\begin{Verbatim}
def table (C : ℕ → ℕ → β) : ℕ → ℕ → β
  | 0, 0 => C 0 0
  | i + 1, 0 => table C i 0 + C (i + 1) 0
  | 0, j + 1 => table C 0 j + C 0 (j + 1)
  | i + 1, j + 1 =>
      min (table C i j) (min (table C i (j + 1)) (table C (i + 1) j)) +
        C (i + 1) (j + 1)
\end{Verbatim}
For a local cost $c:\alpha\to\alpha\to\mathbb{R}$, DTW is the instance $C(i,j)=c(x_i,y_j)$:
\lean{dtw c x y m n = table (localCost c x y) m n}. This is the recurrence of
Keogh and Ratanamahatana~\cite[Eq.~(5)]{keogh2005exact} and of M\"uller~\cite[Thm.~4.3]{muller2007dtw},
with equal weights on the three steps. Keogh and Ratanamahatana's DTW
value~\cite[Eq.~(4)]{keogh2005exact} also takes a square root of the sum. We work with the sum
throughout. Since $\sqrt{\cdot}$ is monotone, each bound $L\le\DTW$ below with $L\ge0$ gives
$\sqrt{L}\le\sqrt{\DTW}$. The square root must be applied to both sides: read literally, their LB\_Kim
reports squared feature differences~\cite[\S3.2]{keogh2005exact}, which is not a lower bound on the
square-rooted value (for $x=(3)$ and $y=(0)$ it gives $9>3$).

\section{The recurrence computes the minimum}\label{sec:dp}

\begin{theorem}[\lean{isLeast\_table}]\label{thm:dp}
Let $\beta$ be a linearly ordered additive commutative monoid and $C:\mathbb{N}\to\mathbb{N}\to\beta$.
Then \lean{table C m n} is the least element of $\{\mathrm{cost}_C(p) : p \text{ a warping path to } (m,n)\}$.
\end{theorem}
\begin{proof}[Proof sketch]
\emph{Lower bound} (\lean{table\_le\_pathCost}). Induct on the path, removing its last cell $q$.
If $r$ is the previous cell, $r\to q$ is a step, and the term $\mathrm{table}(r)$ appears in the
$\min$ defining $\mathrm{table}(q)$. Hence
$\mathrm{table}(q)\le \mathrm{table}(r)+C(q)\le \mathrm{cost}(p')+C(q)$, where $p'$ is the path
without $q$ and the second inequality is the induction hypothesis. The boundary row and column need
a separate case.
\emph{Attainment} (\lean{exists\_pathCost\_eq\_table}). Use functional induction along the equations
of \lean{table}. At $(i{+}1,j{+}1)$ the $\min$ is attained by some predecessor. Extend an optimal path
to it, given by the induction hypothesis, by the cell $(i{+}1,j{+}1)$.
\end{proof}
No nonnegativity or finiteness argument is used: the proof constructs the minimizer. For real costs
we also define the textbook value $\mathrm{dtwSpec} := \inf\{\mathrm{cost}(p)\}$ and obtain
\lean{dtw\_eq\_dtwSpec}, together with the recurrence for \lean{dtwSpec}.

\section{Basic properties and two clarifications}\label{sec:props}

\begin{proposition}
\begin{enumerate}
\item If $c\ge 0$ then $\DTW\ge 0$ (\lean{dtw\_nonneg}).
\item $\DTW_c(x,y)=\DTW_{\mathrm{flip}\,c}(y,x)$ with no assumptions (\lean{dtw\_flip}). Hence, if $c$
  is symmetric, $\DTW_c(x,y)=\DTW_c(y,x)$ (\lean{dtw\_comm}).
\item If $c\ge 0$ and $c(a,a)=0$ then $\DTW(x,x)=0$ (\lean{dtw\_self}).
\end{enumerate}
\end{proposition}
Symmetry follows from the transpose identity $\mathrm{table}_C(i,j)=\mathrm{table}_{C^{\top}}(j,i)$
(\lean{table\_transpose}), where $C^{\top}(i,j)=C(j,i)$. The two straight
predecessors swap roles, which commutativity of $\min$ absorbs. Identity follows from the diagonal:
$\mathrm{table}(i{+}1,i{+}1)\le \mathrm{table}(i,i)+c(x_{i+1},x_{i+1})$.

That DTW is not a metric is well known: M\"uller gives a counterexample for a discrete metric
cost~\cite[Example~4.2]{muller2007dtw}, and Lemire notes the failure and proves a weaker inequality
for banded DTW~\cite{lemire2009faster}. We give a kernel-checked witness.
\begin{theorem}[\lean{not\_triangle\_abs}]
For $c(a,b)=|a-b|$, DTW violates the triangle inequality on real sequences of length~$3$. For
$x=(0,0,0)$, $y=(0,0,1)$ and $z=(0,1,1)$: $\DTW(x,z)=2>1+0=\DTW(x,y)+\DTW(y,z)$.
\end{theorem}
The three values are evaluated from the definition of \lean{dtw} by \lean{norm\_num}; they are not
entered by hand. The witness was found by exhaustive search over sequences with entries in
$\{0,1,2\}$ (\lean{scripts/find\_triangle\_counterexample.py}). The script found no equal-length
witness of length $\le 2$ over that alphabet for $|a-b|$. The same sequences also violate the
inequality for $c(a,b)=(a-b)^2$ (\lean{not\_triangle\_sq}). That case is less informative: the squared
cost is not a metric itself, and the script finds a violation already at length~$1$
($(0),(1),(2)$ give $4>1+1$).

\paragraph{Path conditions.}
Keogh and Ratanamahatana state boundary, continuity ($a-a'\le 1$, $b-b'\le1$) and monotonicity
($a-a'\ge 0$, $b-b'\ge 0$) conditions on consecutive cells~\cite[\S2.1]{keogh2005exact}. Read
literally, these admit a repeated cell; M\"uller's step-size condition excludes
it~\cite[Def.~4.1]{muller2007dtw}. We prove (\lean{step\_iff}) that our step relation is exactly
continuity $\wedge$ monotonicity $\wedge$ $r\neq q$, i.e.\ M\"uller's condition. For nonnegative costs
a repeated cell cannot lower the cost, so the two minima agree; this remark is informal and not
formalized. For signed costs they would differ, which is one reason the generic
Theorem~\ref{thm:dp} uses the stricter relation.

\paragraph{Path length.}
The same source states $\max(m,n)\le K< m+n-1$ for the length $K$ of a warping path between sequences
of lengths $m,n$~\cite[Eq.~(3)]{keogh2005exact}, and so does~\cite[Eq.~(3)]{ratanamahatana2004making}.
We prove the bounds $\max(m,n)+1\le K\le m+n+1$ in our
0-indexed convention (last indices $m,n$), and that \emph{both} are attained
(\lean{exists\_isWarpingPath\_length\_eq}, \lean{exists\_isWarpingPath\_length\_eq\_max}). In the
1-indexed notation the upper bound is therefore $K\le m+n-1$, attained by the path along two edges of
the matrix. This is an off-by-one slip of no consequence for the rest of those papers.

\begin{lemma}[\lean{IsWarpingPath.\{bounds, exists\_mem\_row, exists\_mem\_col, nodup\}}]\label{lem:bounds}
Every cell of a warping path to $(m,n)$ lies in $[0,m]\times[0,n]$. Every row $i\le m$ and every column
$j\le n$ is visited. No cell is visited twice.
\end{lemma}
Row coverage holds because the row index increases by at most one per step. Column coverage follows
by transposing the path. There are no repeats because $i+j$ strictly increases along the path.

\section{Global constraints and the Sakoe--Chiba band}\label{sec:band}

A global constraint is a decidable region $R\subseteq\mathbb{N}^2$. The standard way to compute
constrained DTW is to set the local cost to $\infty$ outside~$R$~\cite[\S4.2.3]{muller2007dtw}. We
make this literal: constrained DTW runs the same DP over $\beta=\mathbb{R}\cup\{\top\}$ (mathlib's
\lean{WithTop ℝ}), with cost $c(x_i,y_j)$ inside $R$ and $\top$ outside. In Lean,
\lean{dtwRegion R c x y m n = table (restrict R (localCost c x y)) m n}.
A path's restricted cost is its real cost if the path stays in $R$, and $\top$ otherwise (\lean{pathCost\_restrict}). So
Theorem~\ref{thm:dp} immediately gives two facts. Constrained DTW is at most the cost of every
admissible path (\lean{dtwRegion\_le\_of\_path}). When it is finite, it is attained by one
(\lean{exists\_path\_of\_dtwRegion\_ne\_top}). Using $\top$ instead of $\inf\emptyset$ matters: in
$\mathbb{R}$, mathlib's convention $\inf\emptyset=0$ would silently make infeasible banded DTW equal to
zero.

The Sakoe--Chiba band of half-width $w$ is $\{(i,j): |i-j|\le w\}$, following the form $j-r\le i\le j+r$
of~\cite[\S3.3.3]{keogh2005exact}. This band is not rescaled for unequal lengths. M\"uller's version
follows the slope $(M-T)/(N-T)$ of the rectangle~\cite[\S4.2.3]{muller2007dtw}, so part~3 below is
specific to the unscaled form.

\begin{theorem}\label{thm:band}
Write $\DTW_w$ for DTW restricted to the band of half-width $w$.
\begin{enumerate}
\item $\DTW\le\DTW_R$ for every region $R$, and $R\subseteq R'$ implies $\DTW_{R'}\le\DTW_R$
  (\lean{dtw\_le\_dtwRegion}, \lean{dtwRegion\_anti}). In particular $\DTW\le\DTW_w$ and
  $w\mapsto\DTW_w$ is antitone (\lean{dtw\_le\_dtwBand}, \lean{dtwBand\_antitone}).
\item If $w\ge\max(m,n)$ then $\DTW_w=\DTW$ (\lean{dtwBand\_eq\_dtw}).
\item $\DTW_w(x,y)<\top$ if and only if $|m-n|\le w$ (\lean{dtwBand\_ne\_top\_iff}).
\end{enumerate}
\end{theorem}
Parts 1 and 2 follow from two facts about the table. First, it is monotone in the cost
(\lean{table\_mono}). Second, $\mathrm{table}(m,n)$ depends only on costs in $[0,m]\times[0,n]$
(\lean{table\_congr}). Part 3 is proved separately, by an induction showing that every in-band cell
other than $(0,0)$ has an in-band predecessor.

\section{Lower bounds}\label{sec:lb}

All bounds use one principle. DTW is attained by some warping path (Theorem~\ref{thm:dp}), so any
inequality that holds for \emph{every} path also holds for DTW (\lean{le\_dtw\_of\_forall\_path}).

\paragraph{LB\_Kim.}
Keogh and Ratanamahatana describe LB\_Kim~\cite{kim2001index} as follows: ``The features are the first
and last elements of the sequence, together with the maximum and minimum values. The maximum squared
differences of corresponding features are reported as the lower bound''~\cite[\S3.2]{keogh2005exact}.
Rakthanmanon et al.\ use the $O(1)$ variant LB\_KimFL, built from the first and last pairs
only~\cite[\S4.1.2]{rakthanmanon2012searching}. They do not say how the two distances are combined;
we add them, which part~1 below shows is valid. We formalize the definitions as described in these two
secondary sources, not from the original papers~\cite{kim2001index,yi1998efficient}.

\begin{theorem}[\lean{lbKimFL\_le\_dtw}, \lean{lbKim\_le\_dtw}]
\begin{enumerate}
\item If $c\ge0$ and $(m,n)\neq(0,0)$, then $c(x_0,y_0)+c(x_m,y_n)\le\DTW_c(x,y)$.
\item Let $c(a,b)=f(|a-b|)$ with $f$ monotone and nonnegative on $[0,\infty)$. Then
$\max\{f|x_0-y_0|,\ f|x_m-y_n|,\ f|\max x-\max y|,\ f|\min x-\min y|\}\le\DTW_c(x,y)$.
\end{enumerate}
\end{theorem}
For part~1, the first and last cells are distinct cells of every path. For part~2, suppose
$\max x\ge\max y$ and $x_{i^*}=\max x$. Row coverage gives a cell $(i^*,j)$ on the path, and
$|\max x-\max y|\le x_{i^*}-y_j\le|x_{i^*}-y_j|$. The other case uses column coverage, and the
minimum feature is symmetric. Each feature is thus bounded by the cost of one cell, and hence by the
whole path. Taking $f(t)=t$ and $f(t)=t^2$ gives the absolute and squared versions
(\lean{lbKim\_le\_dtw\_sq}).

\paragraph{LB\_Keogh.}
Keogh and Ratanamahatana define, for equal lengths and band reach $r$, the envelope
$U_i=\max(q_{i-r}:q_{i+r})$, $L_i=\min(q_{i-r}:q_{i+r})$~\cite[Eqs.~(6)--(7)]{keogh2005exact}, and sum
the squared distances of the other sequence to it~\cite[Eq.~(9)]{keogh2005exact}. Write $d_j(a)$ for
the distance from $a$ to $[L_j,U_j]$, defined as in their Eq.~(9).

The proof below never uses the shape of the band, only that the chosen cells lie in the admissible
region. We therefore state it for an arbitrary global constraint $R$ and sequences of possibly
different lengths. The envelope of column $j$ is taken over the rows that $R$ admits:
$U_j=\max\{x_k: k\le m,\ (k,j)\in R\}$ and $L_j=\min\{\cdots\}$.

\begin{theorem}[\lean{lbKeoghRegion\_le\_dtwRegion}]
Let $R$ be any decidable region and $c(a,b)=f(|a-b|)$ with $f$ monotone and nonnegative on
$[0,\infty)$. Then $\sum_{j=0}^{n} f\big(d_j(y_j)\big)\le\DTW_R(x,y)$.
\end{theorem}
\begin{proof}[Proof sketch]
If $\DTW_R=\top$ there is nothing to prove. Otherwise take an optimal path $p$ inside $R$. For each
column $j$, choose a cell $(g(j),j)\in p$ (column coverage). Since $(g(j),j)\in R$ and $g(j)\le m$, we
have $x_{g(j)}\in[L_j,U_j]$, so $d_j(y_j)\le|x_{g(j)}-y_j|$. The cells $(g(j),j)$ are pairwise
distinct, and $p$ has no repeated cells (Lemma~\ref{lem:bounds}). Hence
$\sum_j C(g(j),j)\le\mathrm{cost}(p)=\DTW_R$.
\end{proof}
Taking $R$ to be the Sakoe--Chiba band, equal lengths and $f(t)=t^2$ gives the constant-width case of
Proposition~1 of~\cite{keogh2005exact}, without the outer square root (\lean{lbKeogh\_le\_dtwBand},
\lean{lbKeogh\_le\_dtwBand\_sq}).

The theorem above is a formally checked restatement, not a new bound. Keogh and Ratanamahatana's
Proposition~1 is stated for sequences of equal length and constraints of the form $j-r\le i\le j+r$;
they apply it with $r$ depending on~$i$, which covers the Itakura
parallelogram~\cite{itakura1975minimum}. Their proof matches each term of the bound with a distinct
cell of the warping path, which is the argument above. Ratanamahatana and Keogh's R-K bands are
arbitrary per-index constraints used together with LB\_Keogh~\cite{ratanamahatana2004making}. Tan et
al.\ remark that their lower-bounding techniques ``trivially generalize to unequal length
series''~\cite{tan2019elastic}. Beyond these, our statement adds only formal generality: the admitted
rows of a column need not form an interval, and $m\neq n$ is allowed, with the envelope taken over the
admitted rows. We do not define the Itakura parallelogram itself in Lean.

\paragraph{LB\_Yi.}
Keogh and Ratanamahatana describe LB\_Yi~\cite{yi1998efficient} as follows: points of one sequence
that are larger (smaller) than the maximum (minimum) of the other ``must contribute at least the
squared difference of their value and the maximum (minimum) value of the other
sequence''~\cite[\S3.2]{keogh2005exact}. This is the instance $R=\mathbb{N}^2$ of the theorem above, where
the envelope degenerates to the global range $[\min x,\max x]$ and constrained DTW is ordinary DTW
(\lean{lbYi\_le\_dtw}, for every cost $f(|a-b|)$ as above).

\section{An executable DP}\label{sec:fast}

\lean{table} is a specification: evaluated naively it is exponential. \lean{dtwFast} implements the
usual algorithm. It keeps one row of $n+1$ values, computes the next row left to right from the
diagonal, upper and left neighbours, and returns entry $n$ of row $m$. We prove
\lean{dtwFast\_eq\_table : dtwFast C m n = table C m n} for every cell cost and every $\beta$, by
showing that each computed row equals the corresponding row of the table.

Over a computable $\beta$ such as $\mathbb{Z}$, the implementation evaluates. Lean's kernel recomputes
the three values of the triangle counterexample by \lean{decide}. As a sanity check we compared
\lean{dtwFast} over $\mathbb{Z}$ with an independent Python implementation on 21 pseudo-random pairs
of integer sequences, including one pair of length-1000 sequences. All 21 values agree
(\lean{scripts/CrossCheck.lean}, \lean{scripts/crosscheck.py}). The $O(mn)$ operation count is an
informal property of the code; it is not proved.

\section{Related work}\label{sec:related}

\paragraph{Formalizations.}
We found no prior formalization of DTW itself. On 2026-11-28 we searched the Isabelle AFP index, the
Lean package registry Reservoir, GitHub code search for ``time warping'' in Lean, Coq/Rocq, Agda,
Dafny and Why3 files, and the web for DTW combined with ``formal verification'', ``mechanized'',
``machine-checked'' and the names of the main proof assistants. We did not search the HOL4, HOL Light
or PVS libraries or opam packages specifically, and GitHub code search covers only default branches.
The record is in \lean{NOVELTY.md} in the repository. The closest work verifies DP algorithms for
related sequence comparisons.
\begin{itemize}
\item Wimmer, Hu and Nipkow verify memoized DP in Isabelle, including minimum edit
  distance~\cite{wimmer2018verified,wimmer2018afp}.
\item The Why3 gallery verifies a linear-space program for an insertion--deletion variant of edit
  distance against an inductive definition of edit distance~\cite{why3editdistance}. A Rocq development proves a recursive
  Levenshtein model minimal over edit scripts, a Wagner--Fischer DP equal to it, and a C
  implementation correct with respect to that DP, using both VST and RefinedC~\cite{joomeditdistance}.
\item Mathlib formerly contained a Levenshtein distance generic in the cost type, computed row by row
  through \lean{suffixLevenshtein} with a proved recurrence~\cite{mathliblevenshtein}. That is the
  same technique as our Section~\ref{sec:fast}, but the file has no theorem relating the value to edit
  scripts. It was removed in 2025~\cite{mathlibrwsearch}.
\item The atlas-lean repository~\cite{atlaslean} proves Needleman--Wunsch~\cite{needleman1970general}
  optimality over alignments in Lean~4. It also proves that the banded alignment optimum (specified by
  enumeration) is at most the unrestricted one, and that a banded alignment exists exactly when the
  length difference is at most the band width, which is the analogue of part~3 of
  Theorem~\ref{thm:band}.
\end{itemize}
Our Theorems~\ref{thm:dp} and~\ref{thm:band} follow the same proof pattern as those alignment
results. What differs is the object. A warping path has no gaps and allows many-to-many matching, and
the lower bounds have no counterpart there. A Rocq/Verus artifact in the liblevenshtein-rust
repository~\cite{liblevenshtein} verifies the arithmetic of a single DTW cell update, an envelope bound
and a band-reachability fact. It also checks a triangle-inequality violation on hand-entered costs.
It does not define DTW over sequences or relate the DP to warping paths.

\paragraph{DTW theory.}
The definitions, the DP theorem, the failure of the triangle inequality and the $\infty$-cost
treatment of global constraints are textbook material~\cite[Ch.~4]{muller2007dtw}. Lemire proves a
weak triangle inequality for banded DTW and introduces LB\_Improved~\cite{lemire2009faster}. Tighter
bounds followed: LB\_Enhanced~\cite{tan2019elastic}, and LB\_Petitjean and
LB\_Webb~\cite{webb2021tight}. Paparrizos et al.\ survey and benchmark lower bounds for DTW and other
elastic measures~\cite{paparrizos2023accelerating}. We formalize none of these newer bounds.

\section{Discussion and limitations}

\begin{itemize}
\item \emph{Univariate lower bounds.} LB\_Kim and LB\_Keogh are proved for real-valued sequences.
  Multivariate costs, as in multichannel EEG, are not covered; per-channel envelopes are the natural
  extension.
\item \emph{Constraints.} Part~1 of Theorem~\ref{thm:band} and LB\_Keogh hold for arbitrary regions;
  parts~2 and~3 are specific to the unscaled Sakoe--Chiba band. Named constraints other than that band
  (e.g.\ the Itakura parallelogram) are not defined in the library.
\item \emph{Step patterns.} The DP adds the cell cost independently of the incoming step. Step
  weights such as the diagonal weight~$2$, other step-size conditions~\cite[\S4.2]{muller2007dtw}
  and subsequence DTW are not covered.
\item \emph{Arithmetic.} All bounds are proved over $\mathbb{R}$, and the executable DP is run over
  $\mathbb{Z}$; floating-point rounding is not modelled.
\item \emph{Complexity.} The row-by-row algorithm is verified for functional correctness only.
\item \emph{Sequences as functions.} Using $\mathbb{N}\to\alpha$ with an explicit last index is
  justified by Lemma~\ref{lem:bounds} and \lean{table\_congr}. It cannot represent empty sequences,
  and we provide no \lean{Fin}- or \lean{List}-indexed interface.
\end{itemize}

\section{Artifact}

The development has 1450 lines of Lean in nine files, counting comments and blank lines, with 105
theorems and lemmas and 36 definitions, structures and instances (\lean{sh scripts/stats.sh}, commit
\lean{77719d9}). It builds against Lean~4.36.0 and
mathlib~v4.36.0. \lean{scripts/Axioms.lean} prints the axioms of 31 main theorems; CI checks that they
are among \lean{propext}, \lean{Classical.choice} and \lean{Quot.sound}, and that the library contains
no \lean{sorry}.
Repository: \url{https://github.com/harimaruthachalam/dtw-lean}.

\paragraph{Use of AI tools.} The Lean development and the text of this paper were drafted with an AI
coding assistant (Claude Code, Anthropic) under the author's direction and review. Every numbered theorem,
lemma and proposition is checked by Lean's kernel, with the axiom audit above; informal remarks in the
prose are marked as such. Every reference was checked against its DOI, publisher, arXiv or repository
page (at the stated commit), and every number was regenerated by the scripts named in the text. The
author takes full responsibility for the content.

\bibliographystyle{plain}
\bibliography{refs}
\end{document}